\documentclass[10pt,letterpaper,twocolumn]{article} 
\usepackage{ol2}
\usepackage[draft]{hyperref}
\usepackage{amsmath}
\usepackage{color}
\usepackage{balance}

\newcommand{\beq}{\begin{equation}}
\newcommand{\eeq}{\end{equation}}
\newcommand{\ket}[1]{|#1\rangle}

\newcommand{\dd}{\text d}
\newcommand{\kk}{{\bf k}}

\begin{document}

\twocolumn[ 

\title{Bi-photon propagation control with optimized wavefront by means of Adaptive Optics}

\author{M. Minozzi,$^1$ S. Bonora,$^2$ A. V. Sergienko,$^3$ G. Vallone,$^{1,2}$ and P. Villoresi$^{1,2}$}

\address{
$^1$Department of Information Engineering, University of Padova, Via Gradenigo, 6/B, 35131, Padova, Italy\\
$^3$Institute for Photonics and Nanotechnology, Nat. Res. Council, Via Trasea, 7, 35131, Padova, Italy\\
$^2$Department of Electrical and Computer Engineering, and Photonics Center, Boston University, 8 Saint Mary's Street, Boston, Massachusetts 02215}

\begin{abstract*}
We present an efficient method to control the spatial modes of entangled photons produced through SPDC process. Bi-photon beam propagation is controlled by a deformable mirror, that shapes a $404nm$ CW diode laser pump interacting with a nonlinear BBO type-I crystal. Using such adaptive optics the propagation the biphoton field  is optimized, exploiting  a feedback signal extracted from the two photon coincidence rate and used  to modify the pump wavefront. We also demonstrated
the enhancement of the biphoton wavefunction coupling into single spatial modes.
\end{abstract*}


] 

 {The importance of pump wavefront on the generation of entangled photons in Spontaneous Parametric Down Conversion (SPDC) process has been theoretically studied in the nineties \cite{2PO,2PGO}. 
Deformable mirrors are key devices for the modification of the pump wavefront allowing a full control of the pump beam.
Recently, deformable mirrors have been used in a several experiments in quantum optics\cite{JOSAB,abou02prl,bona08prl}. 
In this work we experimentally exploit the potentialities of a deformable mirror in the SPDC coupling, 
by realizing an experiment in which we shape and 
control the pump wavefront in order to optimize the divergence of correlated two-photon pairs along $2m$ of free propagation. }

In the SPDC process, a pair of photons (signal and idler) is probabilistically emitted from a nonlinear crystal shined by a pump laser beam.
The main goal of the present manuscript is the demonstration that manipulation of  pump wavefront with a deformable mirror could shape the produced down-converted beams. In particular, we will demonstrate the optimized diffraction in the propagation of one of the two SPDC photons. 
Since down-converted light  retains the properties of of the pump wavefront, the effect of wavefront shaping on the pump beam is induced on the biphoton wavefunction. By considering the pump beam propagating in the $z$ direction, it is possible to show \cite{Mod} that the bi-photon wave function can be written as
\beq
\ket{\psi}\!\!=\!\!\int\!\!\dd^2\kk_s\dd^2\kk_i\dd\omega_s\dd\omega_i
\widetilde\psi(\kk_i,\kk_s,\omega_i,\omega_s)a^\dag_{\kk_s,\omega_s}a^\dag_{\kk_i,\omega_i}\ket{\text{vac}}
\eeq
with
\beq
\widetilde\psi(\kk_i,\kk_s,\omega_i,\omega_s)
=\mathcal N L\;A_p(\kk_s+\kk_i,\omega_s+\omega_i)\;
\text{sinc}\left(\frac L2\Delta k_z\right)
\eeq
In the previous equations $L$ is the crystal length, $\kk_i$ and $\kk_s$ are the transverse momentum coordinates of the signal and idler photons,
$A_p(\kk,\omega)$ is the pump profile in the momentum-frequency space and $\Delta k_z$ is
the (longitudinal) phase mismatch $\Delta k_z=k_{pz}-k_{sz}-k_{iz}$, while the term
 $\mathcal N = \epsilon_0\chi^{(2)} {\it E}_pE_sE_i/(2 i \hbar)$ is a function of $\it E$, the fields strengths. 
 
 \begin{figure}[ttb]
\centerline{\includegraphics[width=0.9\columnwidth]{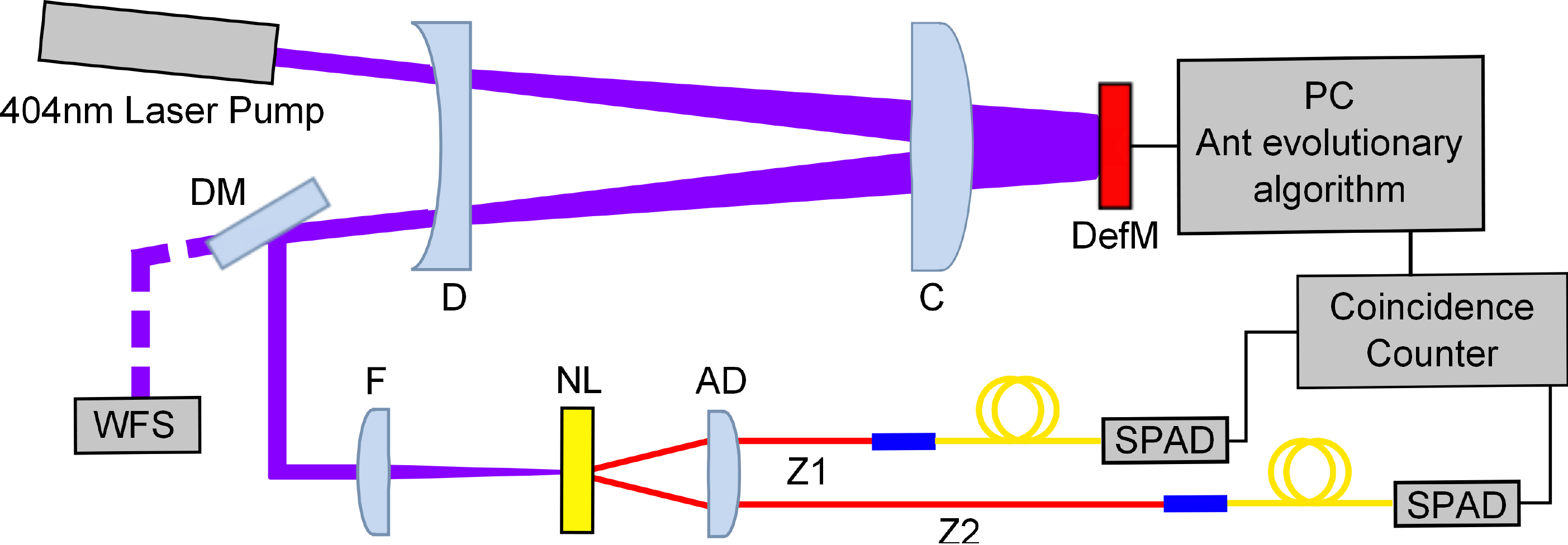}}
\caption{Schematic representation of the experimental setup. The $404nm$ laser passes through a Galilean telescope (lenses \textit{D} and \textit{C}, with $f_D=-250mm$, $f_C=500mm$). After reflection on the deformable mirror ({DefM}), the beam is focused onto a non-linear crystal ({NL}) by the lens  {F} ($f_F=287mm$). A dichroic mirror ({DM}) sends $10\%$ of the pump to a wavefront sensor ({WFS}) to measure its shape. The $808nm$ SPDC photons are collimated by an achromatic doublet {AD} ($f_{AD}=75mm$) and sent to fiber coupling {FC} after $Z_1=0.5m$ and $Z_2=2m$ of free space propagation. Finally, SPDC light is measured by two {SPADs} and coincidences are put in feedback with deformable mirror by Ant evolutionary algorithm.}
\label{fig:ExpSetup}
\end{figure}

From the previous equations it is clear that changes in the pump profile $A_p(\kk,\omega)$ result in 
 changes in the biphoton wave function. Therefore, slight modulations in pump wavefront result in substantial 
 alterations in coincidence counting rate, which can be used as a trace for a well-collimated beam. Consequently, the search for a suitable wavefront, which can produce low divergence SPDC beams, can be achieved by fiber coupling optimization. Such a task can be solved by the use of an evolutionary algorithm, which puts in feedback the action of the mirror with coincidence counting rate.

Let's now describe the setup used in the experiment, shown in Fig. \ref{fig:ExpSetup}. 
In order to generate the SPDC photons we use a $404nm$ laser that pumps a nonlinear $\beta$-barium borate (BBO) type-I crystal (NL). 
The waveform modulation of the pump beam is achieved by the use of a membrane deformable mirror (DefM) \cite{AO}. This device is an electrostatic-type deformable mirror driven by 32 actuators arranged in a honeycomb pattern. The mirror has an aperture of $19mm$, with an active region  width of $11mm$. As a result, in order to fully exploit the mirror capabilities, the pump beam has to match deformable mirror active region. Hence, a beam expander is used to double the pump beam diameter ($2.5mm$ of initial waist). After being reflected, the beam is reduced by a second passage through the beam expander. Also, this reverse path provides the amplification of wavefront spatial frequencies, thus magnifying the mirror action. Then, after the crystal, degenerate SPDC photon pairs at $808nm$ are produced. Each photon is coupled into a multi-mode fiber and sent to high-efficiency single photon avalanche photodiodes (SPADs) in order to measure photon coincidences.

As a starting point, coincidence signal is  manually maximized with a plane pump wavefront, which is imposed by the deformable mirror before the $F$ lens. In addition, in order to study the two-photon beam divergence, one multi-mode fiber is set $2m$ away from the crystal, while the other is put $50cm$ from the crystal and used as a probe. Furthermore, two pin-holes ($2mm$ diameter) are set along the longer path ($Z_2$) at a distance of $1.7m$ one from the other, so as to select a defined flight direction.

\begin{figure}[tb]
\centerline{\includegraphics[width=\columnwidth]{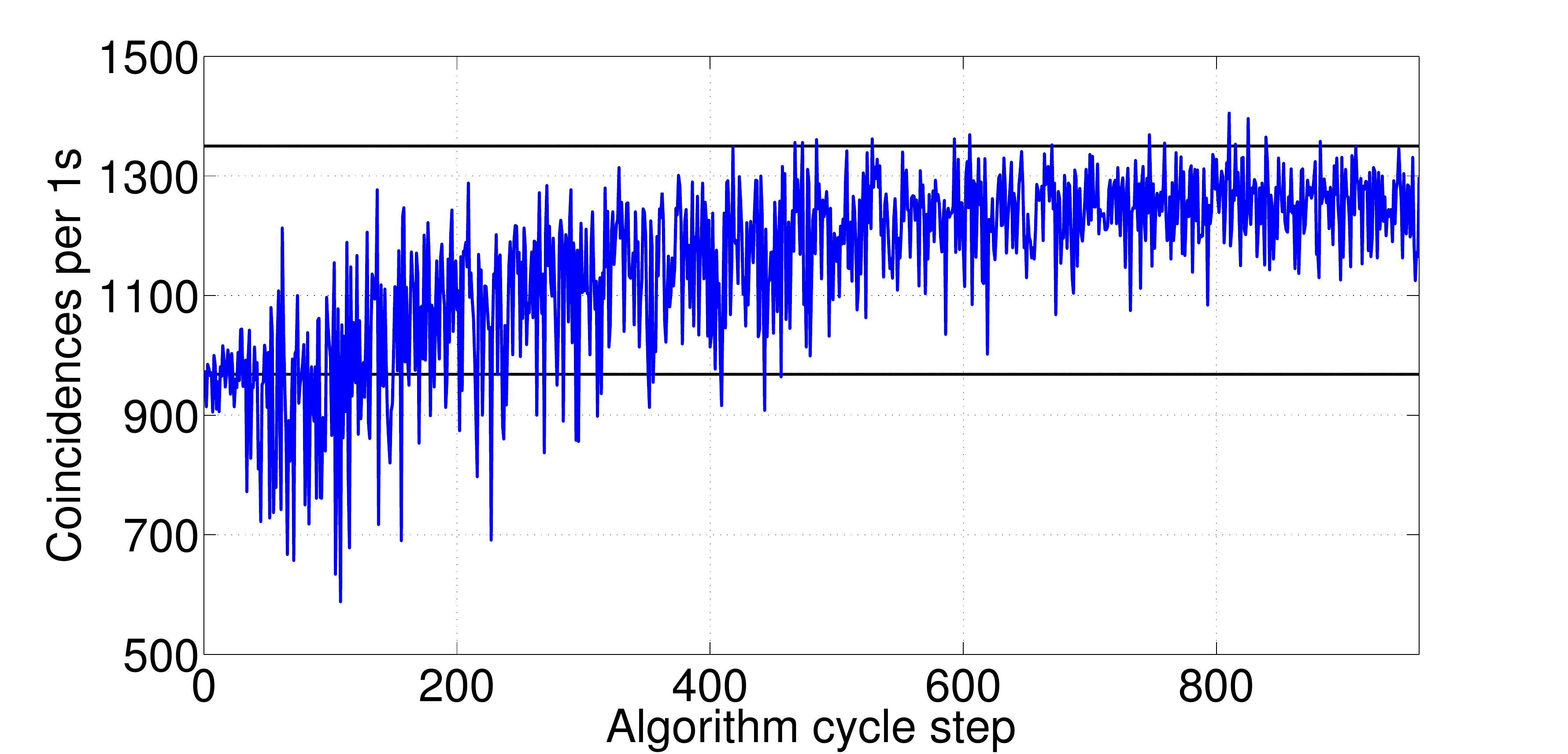}}
\caption{Coincidence evolution provided by the algorithm steps. Ant algorithm maximizes the coincidence counts as explained in the text. 
As a result of the algorithm, a pattern converging to an asymptotic value is achieved. Solid black lines illustrate the average values of initial and final coincidence counts measured over $60s$ of exposure time. 
}
\label{fig:OptAllTry}
\end{figure}
An adaptive algorithm was set up to control SPDC beam divergence over the path defined by the pin-holes. Among the vast choice of such algorithms \cite{AO}, our chiche was the  \textit{Ant colony optimization} \cite{Ant}, because of its small number of free parameters to set, which results in an easier calibration of the algorithm \cite{AO2}. Thus, ant colony optimization is employed to enhance the collection efficiency of SPDC light and, therefore, to generate the  SPDC beam with a ``diffraction-free'' characteristic. In addition, optimization is performed by putting the deformable mirror in feedback with coincidence counts. Thus, mirror actuators are moved randomly by the algorithm and coincidences relative to each actuator configuration are measured. The target is to minimize this following quantity:
\begin{align}
f(\mbox{coinc})=N/\mbox{max}(N,\mbox{coinc})\label{func},
\end{align}
where $N$ stands for a fixed lower bound for coincidence counts, while coincidences measured by SPADs are labeled with \textit{coinc}; in addition $0<f\leq 1$. $N$ is chosen arbitrary by the user, however it is usually set at half of the initial signal. Turning to function (\ref{func}), 
minimizing such a function results in maximizing coincidence counts (this function corresponds to the shortest path in the ``Traveling salesman problem'' \cite{Ant2}).

\begin{figure}[tb]
\centerline{\includegraphics[width=\columnwidth]{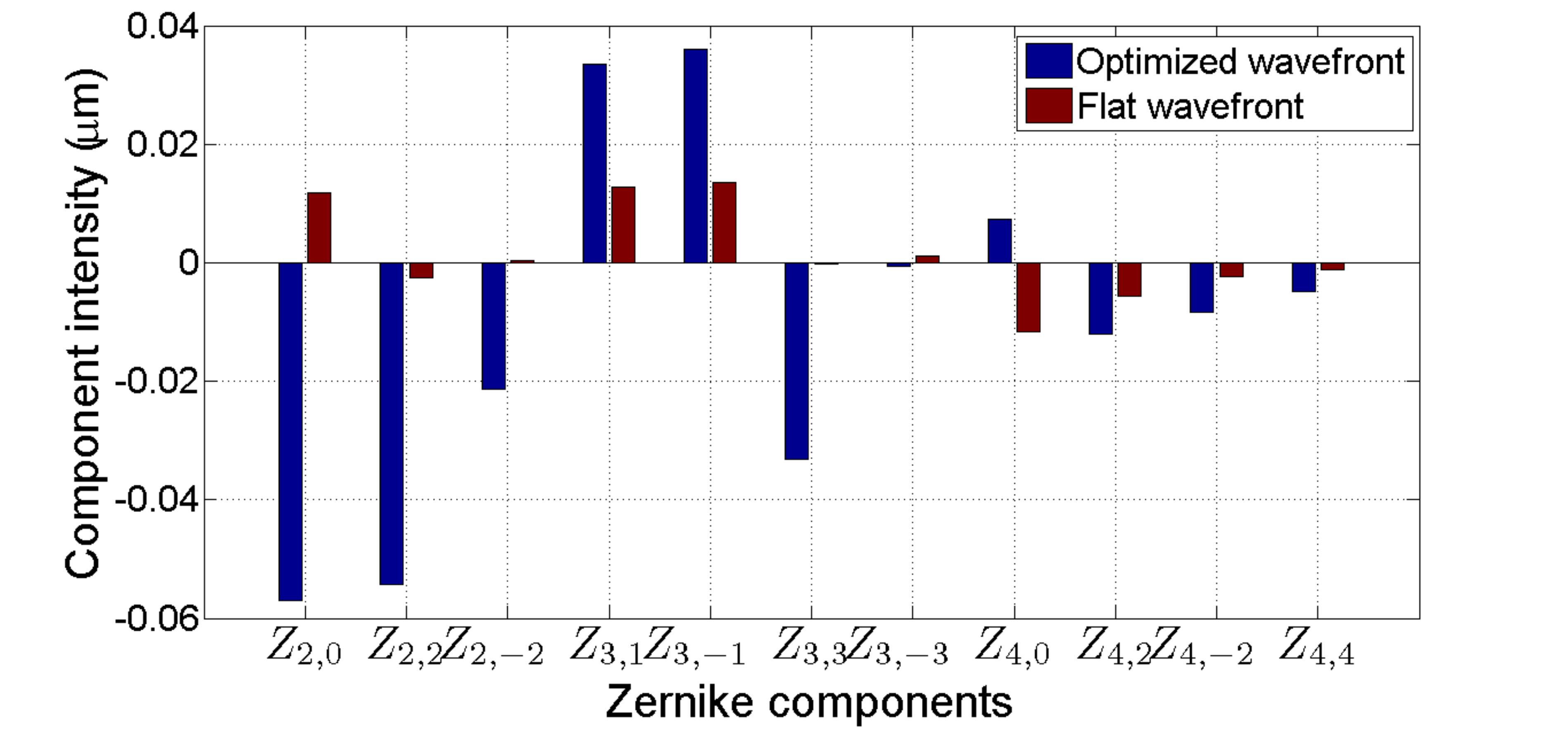}}
\caption{Optimized wavefront compared to initial flat one. The figure illustrates the first four order Zernike components for each wavefront. In the case of flat wavefront ($\mbox{\textit{RMS}}=0.01$), all Zernike terms are less than $0.01\mu m$. Whilst, the optimized wavefront exhibits a significant growth in both second and third order aberrations.}
\label{fig:WavefrontComparison}
\end{figure}
Therefore, minimization is achieved as follows.
Each actuator configuration is associated to a quantity $\tau$ called trial. After setting an initial value of $\tau$ for the first patterns (equal to actuator number), the amount $\Delta\tau\propto 1/f$ is calculated and added to $\tau$, thus updating trial intensity. Then, new configurations are chosen with a probability proportional to $\tau$, thus leading the algorithm to select patterns with the highest amount of trial.
Consequently, mirror configurations which minimize $f$ are achieved. In addition, when actuator configurations provide a value of $f$ less than the previous, this value is stored and mirror configuration are saved. Gradually, the algorithm explores new actuator patterns, refreshing the value of $f$ each time a new minimum is achieved. Over time, the optimum configuration is reached, following an exponential asymptotical growth as shown in Fig. \ref{fig:OptAllTry}.

Turning to initial conditions (initial 32 steps of Fig \ref{fig:OptAllTry}), patterns providing a nearly flat wavefront are selected. This choice is due to the fact that we do not expect significant deviations from flat wavefront, since we measured that strong aberrations ($1\mu m$ of magnitude for each component) result in a loss of coincidence signal. Consequently, we suppose to start from a mirror configuration, which is not so far from coincidence maximum, thus, avoiding any possible \textit{stagnation behavior} \cite{Ant3}. In addition, starting from a ``good position'' allows to speed up algorithm work. For instance, each measurement requires at least $1s$ in order to achieve a signal to noise ratio (\textit{SNR}) less than $5\%$.

\begin{figure}[tb]
\centerline{\includegraphics[width=0.9\columnwidth]{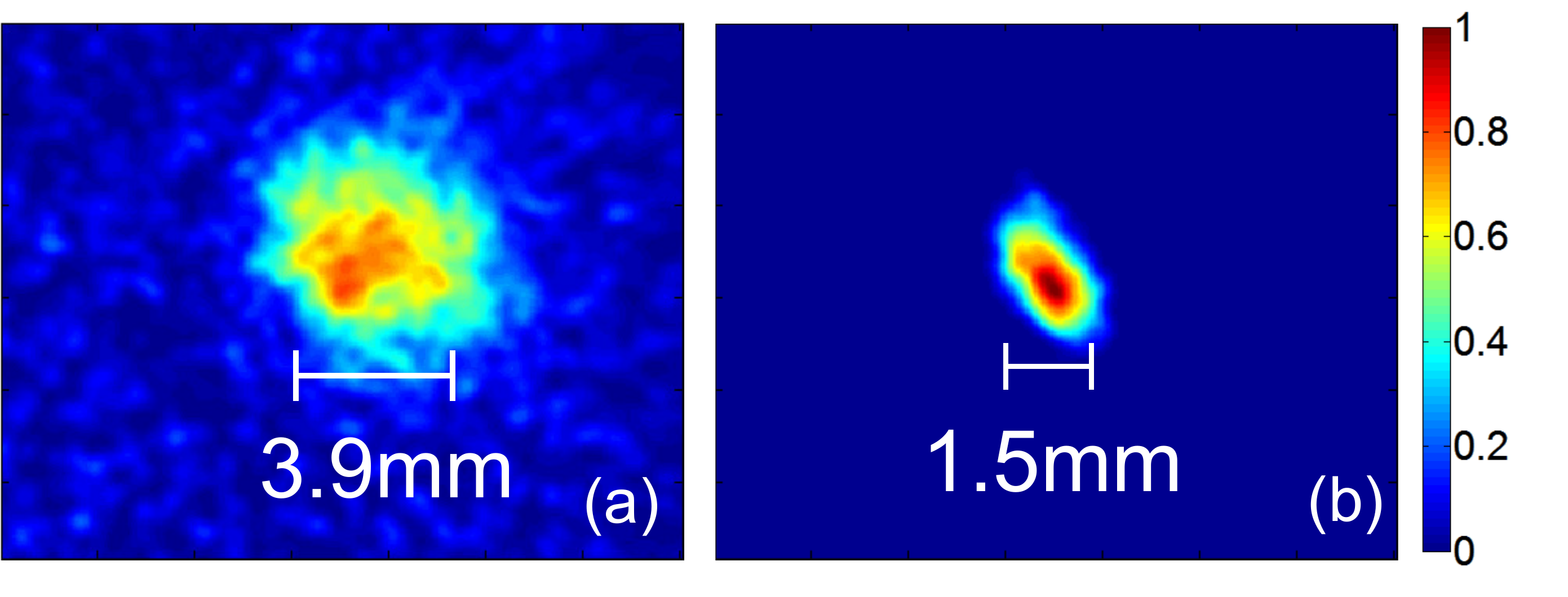}}
\caption{Comparison between SPDC beam spots at $1.5m$ from the crystal. Images are achieved with a flat wavefront (a) and with the optimized one (b). A substantial reduction of beam diameter can be observed.}
\label{fig:Spot}
\end{figure}

Regarding coincidence optimization, several runs of the algorithm showed an increase in coincidences by almost $40\%$. This result means that the system is able to adapt SPDC beam size with the apertures defined by the pin-holes, thus improving the initial signal. This optimization results from a substantial alteration of the pump wavefront, as it is illustrated in Fig. \ref{fig:WavefrontComparison}.
The optimized wavefront is studied in terms of its Zernike coefficient expansion. The most significant one is defocus,  as expected. Nevertheless, fiber coupling optimization is influenced also by higher order terms, such as astigmatism $Z_{2,2}$ and coma (both $Z_{3,1}$ and $Z_{3,-1}$). The overall effect of these aberrations is shown in Fig. \ref{fig:Spot}, which compares SPDC beam spot in front of the second pin-hole (at $1.5m$ from the crystal), before (a) and after (b) the algorithm run.

The system provides a substantial reduction of SPDC spot size, with the optimized spot being almost one fifth on the initial spot. In addition, spot diameter drops from almost $3.9mm$ to about $1.5mm$, due to the action of defocus aberration. As a result, the divergence of SPDC light beam is corrected by the optimization and the beam size is adapted in order to fit pin-hole diameter size. Furthermore, the shape of the spot changes after the algorithm run. For instance, the optimized spot is no more circular, but it is stretched along diagonal direction because of astigmatism and coma aberration. Thus, SPDC beam optimization is the result of both defocus aberration, which compensate SPDC beam divergence, and higher order aberrations (astigmatism and coma), which contribute to beam collimation over long distance.

\begin{figure}[htb]
\centerline{\includegraphics[width=\columnwidth]{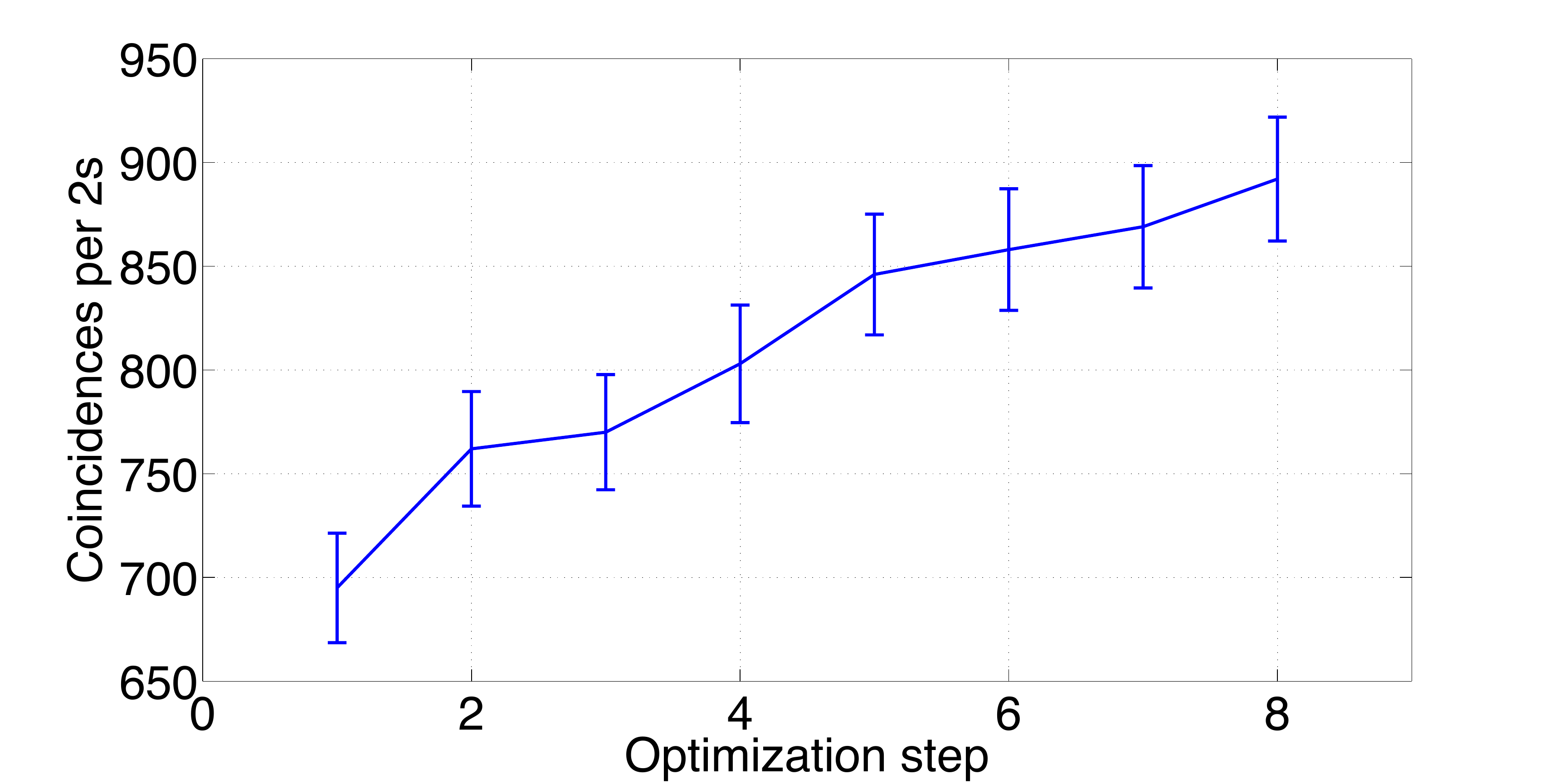}}
\caption{Coincidence evolution provided by algorithm for single mode coupling (we only show the steps increasing the coincidences). }
\label{fig:SM}
\end{figure}
We also tested the algorithm by coupling the radiation into single mode fibers. In Fig. \ref{fig:SM} we show the algorithm steps
in which the coincidence counts are increased. Also in this case we measured the coincidences in 60s before and after the algorithm
optimization. We obtained $20780\pm144$ coincidences at the starting point, while we achieved $25418\pm159$ 
by the optimization algorithm, i.e. the deformation of the mirror allow to increase the coincidences by more than $20\%$.

In conclusion, we have optimized beam divergence of entangled photon pairs so as to obtain ``diffraction-free'' beams over long distances ($2m$). This result is provided by ant evolutionary algorithm which works in feedback with the deformable mirror. These devices are capable to find the best wavefront to center SPDC light beams along a chosen direction and optimize their fiber coupling over $2m$ of free propagation. The algorithm is able to optimize also the coupling into single mode fiber.
Regarding the wavefront imposed by the deformable mirror, it is no more flat, but it exhibits both second and third order aberrations (in addition to a possible tilt), so as to correct bi-photon beam divergence and aberrations introduced by the setup.

The Authors acknowledge the Strategic-Research-Project QUINTET of the Department of
Information Engineering, University of Padova and the
Strategic-Research-Project QUANTUMFUTURE of the University of
Padova.

\end{document}